\documentclass{llncs}

\usepackage[latin9]{inputenc}
\usepackage[T1]{fontenc}

\usepackage{url}
\usepackage{graphicx}
\usepackage[FIGBOTCAP]{subfigure}
\usepackage{multirow}
\usepackage{boxedminipage}
\usepackage{xspace}

\newcommand{\openmp}{OpenMP\xspace}
\newcommand{\marcel}{\emph{Marcel}\xspace}
\newcommand{\bsched}{BubbleSched\xspace}
\newcommand{\magomp}{MaGOMP\xspace}
\newcommand{\libgomp}{\texttt{libgomp}\xspace}
\newcommand{\gomp}{GOMP\xspace}
\newcommand{\gnu}{GNU\xspace}
\newcommand{\gcc}{GCC\xspace}
\newcommand{\posix}{POSIX\xspace}
\newcommand{\upc}{UPC\xspace}

\def\|#1|{\texttt{#1}}

\begin{document}

\title{An Efficient OpenMP Runtime System\\
for Hierarchical Architectures}

\author{
Samuel Thibault \and
François Broquedis \and
Brice Goglin \and \\
Raymond Namyst \and
Pierre-André Wacrenier
}

\institute{
INRIA Futurs - \textsc{LaBRI}\\
351 cours de la libération\\
33405 Talence cedex, France\\
\email{\{thibault,goglin,namyst,wacrenier\}@labri.fr,
francois.broquedis@etu.u-bordeaux1.fr}
}

\maketitle

\begin{abstract}

  Exploiting the full computational power of always deeper
  hierarchical multiprocessor machines requires a very careful
  distribution of threads and data among the underlying non-uniform
  architecture.
  The emergence of multi-core chips and NUMA machines makes it
  important to minimize the number of remote memory accesses,
  to favor cache affinities, and to guarantee fast completion
  of synchronization steps.
  By using the \bsched platform as a threading backend for the \gomp \openmp
  compiler, we are able to easily transpose affinities of thread teams
  into scheduling hints using abstractions called bubbles.
  We then propose a scheduling strategy suited to nested \openmp
  parallelism. The resulting preliminary performance evaluations show an
  important improvement of the speedup on a typical NAS OpenMP benchmark
  application.

  \smallskip
  \noindent\textbf{Keywords:}
  \emph{
    OpenMP, Nested Parallelism, Hierarchical Thread Scheduling,
    Bubbles, Multi-Core, NUMA, SMP.
  }
\end{abstract}

\section{Introduction}

  The emergence of deeply hierarchical architectures based on
  multi-threaded multi-core chips and NUMA machines raises the need
  for a careful distribution of threads and data.
  Indeed, cache misses and NUMA penalties become more and more
  important with the complexity of the machine, making these
  constraints as important as parallelization.
  They require some new programming models and new tools to make
  the most out of these underlying architectures.

  As quoted by Gao \emph{et al.}~\cite{HTVM}, it is important to expose
  domain-specific knowledge semantics to the various software
  components in order to organize computation according to the
  application and architecture.
  Indeed, the whole software stack, from the application to the
  scheduler, should be involved in the parallelizing, scheduling and
  locality adaptation decisions by providing useful information to the
  other components.

  Therefore, in OpenMP frameworks, the information extracted by the
  compiler (about memory affinity and adherence to the same parallel
  section) can be very useful for the guidance of task/thread
  scheduling.
  On the other hand, it is very important to rely on architecture specific
  constraints when making these scheduling decisions.
  A tight interaction between the OpenMP stack and the underlying
  hardware-aware scheduler is thus required.

  The most delicate point, when dealing with irregular applications,
  is to exploit this knowledge at runtime (during the whole execution
  time) so as to maintain a good balancing of threads when events
  arise (task termination, creation of new embedded parallel sections,
  blocking synchronization, etc.).

  In this paper, we propose a hierarchical threading library able to follow/obey
  sched\-uling directives and advices in a very powerful manner. Scheduling
  information (affinity, group membership) is attached to bubbles,
  which are abstractions that can recursively group threads or bubbles
  sharing common properties.

  We report on preliminary experiences on top of a 8-way multi-core
  NUMA machine and we show that running OpenMP applications on top of our
  runtime system greatly enhances performance on hierarchical
  architectures under irregular conditions. We also propose insights
  regarding the extraction of useful information by the compiler for
  our runtime and discuss the addition of a couple of non-standard OpenMP
  directives that would improve performance.

\section{Scheduling Applications Featuring Nested, Irregular Parallelism}

Achieving the best possible performance when programming OpenMP
applications requires developers to expose the parallelism and
to explicitly design their code to drive its parallel behavior.
Therefore, it is quite common nowadays to define per-thread specific
data structures (in order to avoid false-sharing) and use a static,
possibly pre-calculated, distribution of the workload to get good data
locality~\cite{DynPageMigUser}. Indeed, this model suits very well
regular applications with coarse-grain parallelism.

However, this approach is hardly usable when dealing with
irregular applications that rather need a dynamic load balancing
mechanism. The use of complex synchronization schemes, or even
blocking systems calls, may also be responsible for introducing
irregularities regarding the computing load on the available
processors. Using \openmp dynamic scheduling directives can sometimes
improve performance. In some cases, however, it may penalize data
locality or even introduce false sharing effects, which can severely
impact performance on hierarchical architectures.

Another approach is to increase the number of potential parallel tasks
using nested parallelism, so that threads can be dynamically
(re)allocated according to the workload disparity. The performance of
such a dynamic thread management, when supported\footnote{Nested
  parallelism is currently an optional feature in \openmp.}, heavily
relies on the underlying runtime implementation, but also on the
underlying operating system's scheduler. This explains why \openmp
users have been experiencing poor performance with the nested
capabilities of some \openmp compilers, and have ended up performing
explicit thread programming on top of
\openmp~\cite{blikberg05load,OpenMPThreadGroups} or explicitely binding
thread groups to processors~\cite{OpenMPThreadMapGroup}.

Nevertheless, there exists some very good implementations of \openmp
nested parallelism, such as Omni/ST~\cite{tanaka00performance} for
instance. Such implementations are typically based on a fine-grain
thread management system that uses a fixed number of threads to
execute an arbitrary number of \emph{filaments}, as done in the Cilk
multithreaded system~\cite{cilk5-impl}. The performance obtained over
symmetrical multiprocessors is often very good, mostly because many
tasks can be executed sequentially with almost no overhead when all
processors are busy. However, since these systems provide no support
for attaching high level information such as memory affinity to the
generated tasks, many applications will actually achieve poor
performance on hierarchical, NUMA multiprocessors.

One could probably enhance these \openmp implementations to use
affinity information extracted by the compiler so as to better
distribute tasks or threads over the underlying processors. However,
since only the underlying thread scheduler has complete control over
scheduling events such as processor idleness, blocking syscall or even
thread preemption, this information could only be used to influence
task allocation at the beginning of each parallel section.

We believe that a better solution would be to transmit information
extracted by the compiler to the underlying thread scheduler \emph{in
  a persistent manner}, and that only a tight integration of
application-provided meta-data and architecture description can let the
underlying scheduler take appropriate decisions during the whole
application run time. In other words, one can see this configurable
scheduler framework as a domain-specific language enabling scientists
to transfer their knowledge to the runtime system~\cite{HTVM}.

\section{\magomp: an Implementation of GNU OpenMP for Hierarchical Machines}

To evaluate the potential gain of providing a thread scheduler with
persistent information extracted by an \openmp compiler, we have
extended the GNU \openmp runtime system (i.e. the \libgomp library) so
as to rely on the \marcel thread library.  This library provides
facilities for attaching various information to groups of threads,
together with a framework that helps to develop schedulers capable of
using these metadata. Scheduling policies are simply developed as
\emph{plug-ins}.

Before describing our extensions to the \gnu \openmp
compiler suite, we first present the most important features of
the \marcel library.

\subsection{The \emph{Bubble} Scheduling Model}

\begin{figure}[htb]
\centering
\includegraphics[scale=0.7]{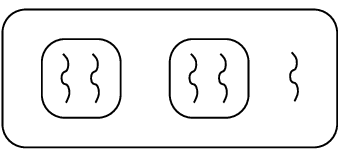}
\hspace{2cm}
\includegraphics[scale=0.7]{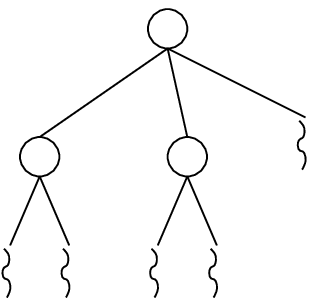}
\caption{Expressing thread relationships: graphical and tree-based
representations.}
\label{ex_bubbles}
\end{figure}

\begin{figure}[htb]
\centering
\includegraphics[scale=0.7]{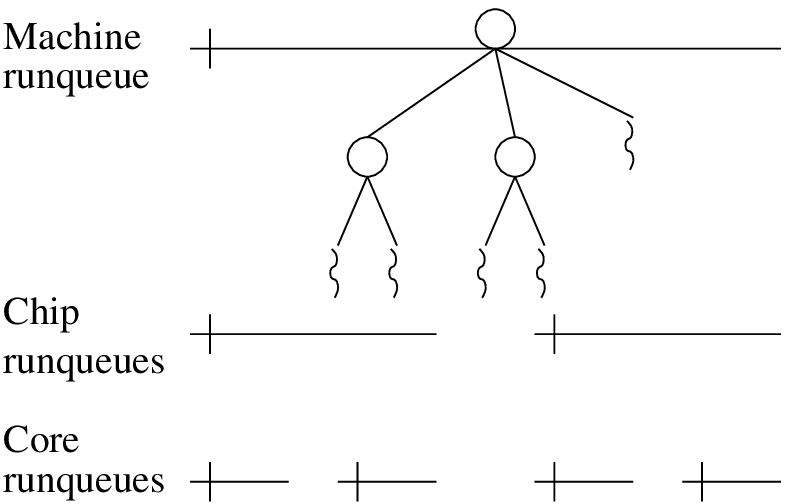}
\hspace{2cm}
\includegraphics[scale=0.7]{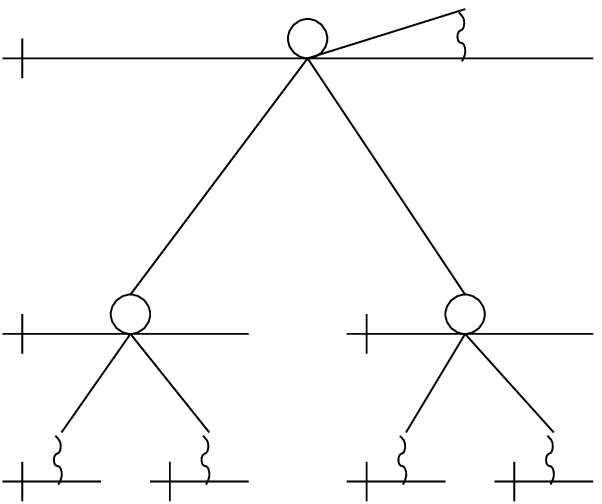}
\caption{Scheduling of bubbles and threads on the runqueues of
a hierarchical machine.}
\label{bubbles_runqueues}
\end{figure}

\marcel is a POSIX-compliant thread library featuring extensions for
easily writing efficient, customized schedulers for hierarchical
architectures. The API of \marcel provides functions to group threads
using nested sets called \textbf{bubbles}~\cite{THIBAULT:2005:31780}.
These abstractions allow programmers to model the relationships
between the different threads of an application. Figure~\ref{ex_bubbles} illustrates
this concept: four threads are grouped as pairs in bubbles (assuming
they work on the same data), which are themselves grouped along
another thread in a larger bubble (assuming they share information
less often). Bubbles allow expression of relationships like data sharing, collective
operations, or more generally a particular scheduling policy need
(serialization, gang scheduling, etc.). Hierarchical machines are
modelled with a hierarchy of runqueues.  Each component of each
hierarchical level of the machine is represented by one runqueue: one
per logical processor, one per core, one per chip, one per NUMA node,
and one for the whole machine. \marcel's ground scheduler then uses a
hierarchical \emph{Self-Scheduling} algorithm. Whenever idle, a
processor scans all runqueues that span it, and executes the first
thread that is found, from bottom to top. For instance, if the thread
is on a runqueue that represents a chip, it may be run by any
processor of this chip (see Figure~\ref{bubbles_runqueues}).

\label{API}
As mentioned previously, \marcel provides a high-level API for
writing powerful and portable schedulers that manipulate threads,
bubbles and runqueues. Threads and bubbles are equally considered as
\textbf{entities}, while bubbles and runqueues are equally considered
as \textbf{scheduling holders}, so that we end up with entities
(threads or bubbles) that we can schedule on holders (bubbles or
runqueues). Primitives are then provided for manipulating entities in
holders.  Runqueues can be accessed through vectors, and can be walked
through thanks to ``parent'' and ``child'' pointers.  Some functions permit to
gather statistics about bubbles so as to take appropriate
decisions. This includes for instance the total number of threads and
the number of running threads, but also various information such as
the accumulated expected and current CPU computation time or memory
usage, or the cache miss rates.

Writing a high-level scheduler actually reduces to writing some hook
functions. The main one is actually called when the ground
Self-Scheduler encounters a bubble during its search for the next
thread to execute. The default implementation just looks for a thread
in the bubble (or one of its sub-bubbles) and switches to it.  The
\verb+bubble_tick()+ hook is called when some time-slice for a bubble
expires, and hence permits periodic operations on bubbles with a
per-bubble notion of time.  Of course, mere ``daemon'' threads can
also be started for performing background operations. As a result,
scheduling experts may manipulate threads with a high level of abstraction by
deciding the placement of bubbles on runqueues, or even temporarily
putting some bubbles aside (by defining their own runqueues that the
basic Self-Scheduler will not look at).

\subsection{Generating Bubbles Out of \openmp Parallel Sections}

The GNU \openmp compiler\cite{gomp}, \gomp, is based on an extension
of the \gcc 4.2 compiler that converts \openmp pragmas into
threading calls.
The creation of threads and teams is actually delegated to a shared
library, \libgomp, which contains an abstraction layer to map \openmp
threads onto various thread implementations.
This way, any application previously compiled by \gomp may be relinked
against an implementation of \libgomp on another thread type and
transparently work the same.

We used this flexible design to develop \magomp, a port of \gomp on
top of the \marcel threading library in which \bsched is implemented.
To do so, a \marcel adaptation of \libgomp threads has been added
to the existing abstraction layer.
We rely on \marcel's fully \posix compatible interface to guarantee that
\magomp will behave as well as \gomp on pthreads.
Then, it becomes possible to run any existing \openmp application on
top of \bsched by simply relinking it.

Once \marcel threads are created they basically behave by default as
native pthreads without any notion of team or memory affinity.
\bsched hooks have been added in the \libgomp code to provide information
about thread teams by creating bubbles accordingly.

Therefore, when a thread encounters a nested parallel region and becomes
the master of a new team, it creates a bubble within its currently holding
bubble.
Then, it moves itself into this new bubble and creates the team's slave
threads inside it.
Finally, the master dispatches the workload across the team.
Once their work is completed, slave threads die while the master destroys
the bubble and returns to its original team.
As shown on Figure~\ref{teamcode}, only a few lines of code are needed
to associate a nested team hierarchy with a bubble hierarchy.

\begin{figure}[ht]
\centering
\begin{boxedminipage}{12cm}
\small
\begin{verbatim}
void gomp_team_start (void (*fn) (void *), void *data, unsigned nthreads,
                      struct gomp_work_share *work_share) {
  struct gomp_team *team;
  team = new_team (nthreads, work_share);
  ... /* Pack 'fn' and 'data' into the 'start_data' structure */

  if (nthreads > 1 && team->prev_ts.team != NULL) {
    /* nested parallelism, insert a marcel bubble */
    marcel_bubble_t *holder = marcel_bubble_holding_task (thr->tid);
    marcel_bubble_init (&team->bubble);
    marcel_bubble_insertbubble (holder, &team->bubble);
    marcel_bubble_inserttask (&team->bubble, thr->tid);
    marcel_attr_setinitbubble (&gomp_thread_attr, &team->bubble);
  }

  for(int i=1; i < nbthreads; i++) {
      pthread_create (NULL, &gomp_thread_attr,
                      gomp_thread_start, start_data);
      ...
  }
}
\end{verbatim}
\end{boxedminipage}
\caption{
  One-to-One correspondence between \marcel's bubble and \gomp's
  team hierarchies.
}
\label{teamcode}
\end{figure}

\subsection{A Scheduling Strategy Suited to \openmp Nested Parallelism}

     The challenge of a scheduler for the nested parallelism of \openmp
resides in how to distribute the threads over the machine. This must be
done in a way that favors both a good balancing of the computation and, in
the case of multi-core and NUMA machines, a good affinity of threads, for
better cache effects and avoiding the remote memory access penalty.

     For achieving this, we wrote a \textbf{bubble spread} scheduler
consisting of a mere recursive function that uses the API described in
section~\ref{API} to greedily distribute the
hierarchy of bubbles and threads over the hierarchy of runqueues. This
function takes in an array of ``current entities'' and an array of
``current runqueues''. It first sorts the list of current entities
according to their computation load (either explicitly specified by the
programmer, or inferred from the number of threads). It then greedily
distributes them onto the current runqueues by keeping assigning the
biggest entity to the least loaded runqueue\footnote{This algorithm
comes from the greedy algorithm typically used for resolving the
bi-partition problem.}, and recurse separately into the sub-runqueues of
each current runqueue.

     It often happens that an entity is much more loaded than others
(because it is a very deep hierarchical bubble for instance). In such
a case, a recursive call is made with this bubble ``exploded'': the
bubble is removed from the ``current entities'' and replaced by its content
(bubbles and threads).  How big a bubble needs to be for being exploded
is a parameter that has to be tuned. This may depend on the
application itself, since it permits to choose between respecting
affinities (by pulling intact bubbles as low as possible) and balancing
the computation load (by exploding bubbles for having small entities for
better distribution).

     This way, affinities between threads are taken into account: since
they are by construction in the same bubble hierarchy, the threads of
the same external loop iterations are spread together on the same NUMA
node or the same multicore chip for instance, thus reducing the NUMA
penalty and enhancing cache effects.

     Other repartition algorithms are of course possible, we are
currently working on a even more affinity-based algorithm that avoids
bubble explosions as much as possible.

\section{Performance Evaluation}

     We validated our approach by experimenting with the BT-MZ
application.  It is one of the 3D Fluid-Dynamics simulation
applications of the Multi-Zone version of the NAS Parallel
Benchmark~\cite{NPB-MZ} 3.2.  In this version, the mesh is split in the
$x$ and $y$ directions into zones.  Parallelization is then performed
twice: simulation can be performed rather independently on the different
zones with periodic face data exchange (coarse grain \emph{outer}
parallelization), and simulation itself can be parallelized among the
$z$ axis (fine grain \emph{inner} parallelization).  As opposed to other
Multi-Zone NAS Parallel Benchmarks, the BT-MZ case is
interesting because zones have very irregular sizes: the size of the
biggest zone can be as big as 25 times the size of the smallest one.  In
the original SMP source
code, outer parallelization is achieved by using Unix processes while
the inner parallelization is achieved through an \openmp static parallel
section. Similarly to Ayguade \emph{et al.}~\cite{NestedFluid}, we modified this to use
two nested \openmp static parallel sections instead, using $n_o*n_i$
threads.

     The target machine holds 8 dual-core AMD Opteron 1.8GHz NUMA chips (hence
a total of 16 cores) and 64GB of memory. The measured NUMA factor
between chips\footnote{The NUMA factor is the ratio between remote memory access
and local memory access times.} varies from 1.06 (for neighbor chips) to 1.4
(for most distant chips). We used the class A problem, composed of
16 zones. We tested both the Native POSIX Thread Library of Linux 2.6
(NPTL) and the \marcel library, before trying the \marcel library with
our \emph{bubble spread} scheduler.

     We first tried non-nested approaches by only enabling either outer
parallelism or inner parallelism, as shown in Figure~\ref{bt-mz}:

\begin{figure}[tb]
\center
\includegraphics[scale=.8]{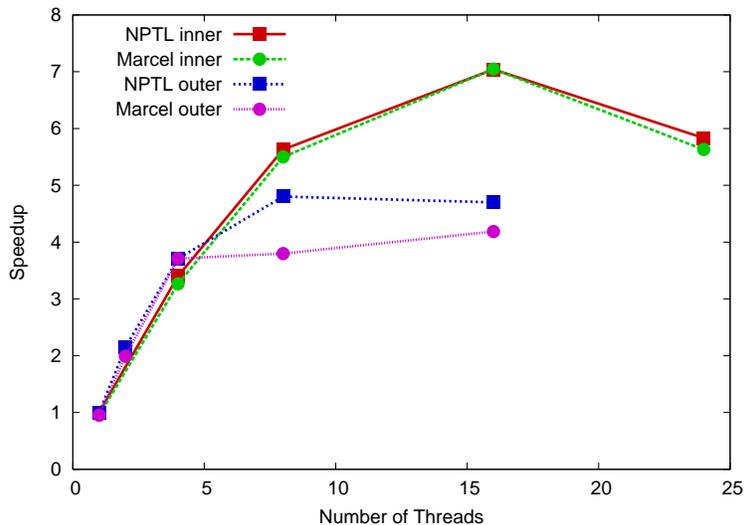}
\caption{\label{bt-mz}Outer parallelism ($n_o*1$) and inner
parallelism ($1*n_i$).}
\end{figure}

\begin{description}
\item [Outer parallelism($n_o*1$):] Zones themselves are distributed among the
processors. Due to the irregular sizes of zones and the fact that there
is only a few of them, the computation is not well balanced, and hence
the achieved speedup is limited by the biggest zones.
\item [Inner parallelism($1*n_i$):] Simulation in zones are performed
sequentially, but simulations themselves are parallelized among
the $z$ axis. The computation balance is excellent, but the nature
of the simulation introduces a lot of inter-processor data exchange.
Particularly because of the NUMA nature of the machine, the speedup is
hence limited to 7.
\end{description}

\begin{figure}
\centering
\subfigure[NPTL]   {\includegraphics[scale=0.6]{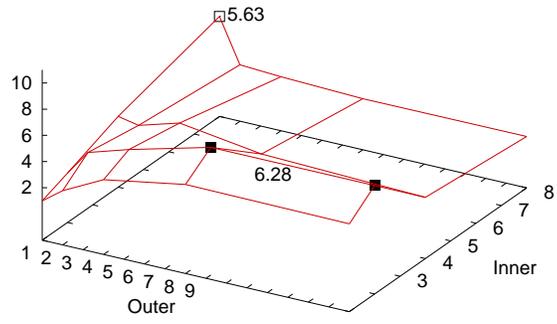}}
\subfigure[Marcel] {\includegraphics[scale=0.6]{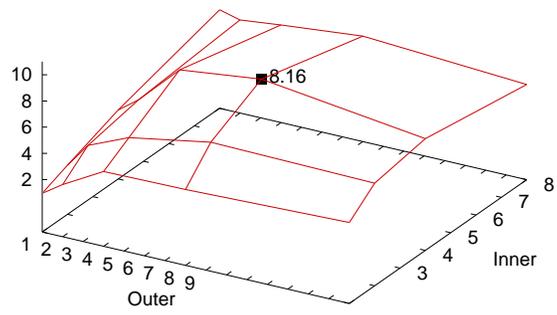}}
\subfigure[Bubbles]{\includegraphics[scale=0.6]{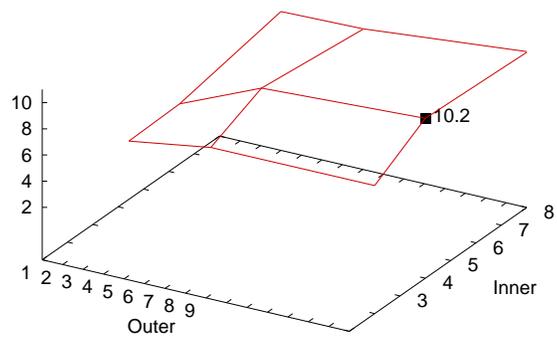}}
\caption{\label{bt-mz-nested}Nested parallelism.}
\end{figure}

    So as to get the benefits of both approaches (locality and balance),
we then tried the nested approach by enabling both parallelisms.  As
discussed by \textsc{Duran} \emph{et al.}~\cite{ThDistrOpenMP}, the
achieved speedup depends on the relative number of threads created by
the inner and the outer parallelisms, so we tried up to 16 threads for
the outer parallelism (i.e. the maximum since there are 16 zones),
and up to 8 threads for the inner parallelism.  The results are shown
on Figure~\ref{bt-mz-nested}.  The nested speedup achieved by NPTL is
very limited (up to $6.28)$, and is actually worse than what pure inner
parallelism can achieve (almost $7$, not represented here because the
"Inner" axis maximum was truncated to 8 for better readability).
\marcel behaves better (probably because user threads are more
lightweight), but it still can not achieve a better speedup than
$8.16$.  This is due to the fact that neither NPTL nor \marcel takes
affinities of threads into account, leading to very frequent remote
memory accesses, cache invalidation, etc.  We hence used our bubble
strategy to distribute the bubble hierarchy corresponding to the
nested \openmp parallelism over the whole machine, and could then
achieve better results (up to $10.2$ speedup with $16*4$ threads). This
improvement is due to the fact that the bubble strategy carefully
distribute the computation over the machine (on runqueues) in an
affinity-aware way (the bubble hierarchy).

    It must be noted that for achieving the latter result, the only
addition we had to do to the BT-MZ source code is the following line:
\small
\begin{verbatim}
call marcel_set_load(int(proc_zone_size(myid+1)))
\end{verbatim}
\normalsize
that explicitly tells the bubble spread scheduler the load of each zone,
so that they can be properly distributed over the machine.  Such a clue
(which could even be dynamic) is very precious for permitting the runtime
environment to make appropriate decisions, and should probably be added
as an extension to the \openmp standard.  Another way to achieve load
balancing would be to create more or less threads according to the
zone size~\cite{NestedFluid}. This is however a bit more difficult to
implement than the mere function call above.

\section{Conclusion}

In this paper, we discussed the importance of establishing a
persistent cooperation between an \openmp compiler and the underlying
runtime system for achieving high performance on nowadays multi-core
NUMA machines. We showed how we extended the \gnu \openmp
implementation, \gomp, for making use of the flexible \marcel thread
library and its high-level \emph{bubble} abstraction. This permitted
us to implement a scheduling strategy that is suited to \openmp nested
parallelism. The preliminary results show that it improves the
achieved speedup a lot.

At this point, we are enhancing our implementation so as to introduce
just-in-time allocation for \marcel threads, bringing in the notion
of ``ghost'' threads, that would only be allocated when first run by
a processor. In the short term, we will keep validating the obtained
results over several other \openmp applications, such as Ondes3D (French
Atomic Energy Commission). We will compare the resulting performance
with other \openmp compilers and runtimes. We also
intend to develop an extension to the \openmp
standard that will provide programmers with the ability to
specify load information in their applications, which the runtime will
be able to use to efficiently distribute threads.

In the longer run, we plan to
extract the properties of memory affinity at the compiler level, and
express them by injecting gathered information into more accurate
attributes within the bubble abstraction. These properties may be
obtained either thanks to new directives \emph{à la} \upc\footnote{
  The \upc \texttt{forall} statement adds to the traditional
  \texttt{for} statement a fourth field that describes the affinity
  under which to execute the loop
}~\cite{upc} or be computed automatically via static
analysis~\cite{CompilerRefAffinity}.  For instance, this kind
of information is helpful for a bubble-spreading scheduler, as we want
to determine which bubbles to explode or to decide whether or not it is
interesting to apply a \emph{migrate-on-next-touch}
mecanism~\cite{OpenMPPDENUMA} upon a scheduler decision.  
All these extensions will rely on a memory management library that
attaches information to bubbles according to memory affinity, so that,
when migrating bubbles, the runtime system can migrate not only
threads but also the corresponding data.

\clearpage
\section{Software Availability}

\marcel and \bsched are available for download within the PM2
distribution at
\url{http://runtime.futurs.inria.fr/Runtime/logiciels.html} under the
GPL license.
The \magomp port of \libgomp will be available soon and may be
obtained on demand in the meantime.

\small
\bibliographystyle{alpha}

\bibliography{bib}

\end{document}